\documentclass[journal]{IEEEtran}
\usepackage{cite}
\usepackage{amsmath,amssymb,amsthm,bm}
\usepackage{graphicx}
\usepackage{xcolor}
\usepackage{enumitem}
\usepackage[ruled,linesnumbered]{algorithm2e}
\usepackage[hidelinks]{hyperref}
\newtheorem{proposition}{Proposition}
\newcommand{\R}{\mathbb{R}}
\newcommand{\prox}{\operatorname{prox}}
\newcommand{\tr}{\operatorname{tr}}
\newcommand{\diag}{\operatorname{diag}}
\newcommand{\x}{\bm{x}}\newcommand{\y}{\bm{y}}\newcommand{\rr}{\bm{r}}\newcommand{\A}{\bm{A}}\newcommand{\w}{\bm{w}}
\newcommand{\I}{\bm{I}}\newcommand{\T}{\bm{T}}\newcommand{\Dm}{\bm{D}}\newcommand{\iL}{\bm{i}_L}
\newcommand{\dr}{\delta\bm{r}}\newcommand{\xs}{\x^\star}

\begin{document}

\title{Stabilizing Vector Approximate Message Passing with Circuit-Designed Lag-Lead Damping}

\author{Mohamed Akrout, \IEEEmembership{Member, IEEE}
\thanks{M. Akrout is with the Department of Electrical Engineering and Computer Science, University of Tennessee, Knoxville, TN 37996, USA (e-mail: makrout@tennessee.edu).}}

\maketitle

\begin{abstract}
Vector approximate message passing (VAMP) often fails to converge on ill-conditioned matrices, and its damping factor is tuned either manually or adaptively. Using the fact that VAMP iterations are the discretization of an electric circuit, we show that the denoiser and the likelihood act as nonlinear resistors, the Onsager correction is a mirror resistor and damping is a lag filter. This circuit gives a closed-form loop gain whose spectrum can predict when VAMP fails. Because a lag filter cannot isolate the oscillating modes, we add a lead capacitor whose design often results in a negative capacitance. Across a wide range of condition numbers, our proposed damping strategy achieves a higher convergence rate than adaptive damping and Anderson acceleration at a lower mean iteration cost. On ill-conditioned multipath channels, across ranks and sparsity levels, it also converges more reliably than existing damping baselines.
\end{abstract}


\section{Introduction}
\IEEEPARstart{V}{ector} approximate message passing (VAMP) is among the most effective iterative estimators for the linear inverse problem $\y = \A\x + \w$ \cite{rangan2019vamp}. Its state evolution is exact for rotationally invariant $\A$ in the large-system limit \cite{rangan2019vamp,takeuchi2020ep}. On a fixed ill-conditioned matrix, however, it often oscillates. The usual heuristic solution is damping with a factor chosen by trial and error \cite{rangan2019vamp,rangan2019convergence}. Adaptive damping adjusts this factor over iterations based on a free-energy cost \cite{vila2015damping}, while Anderson acceleration serves as the generic fixed-point accelerator \cite{walker2011anderson}. Long-memory variants of approximate message passing (AMP) change the state evolution rather than the loop \cite{donoho2009amp,rangan2011gamp,ma2017oamp,liu2022mamp}. None of these works, however, explains why VAMP oscillates.

First-order optimization algorithms are discretizations of circuits of resistors, inductors and capacitors, in which the subdifferential of the objective is a nonlinear resistor \cite{boyd2024circuits}. From this perspective, algorithmic design becomes network design. This view builds on continuous-time models of optimization \cite{su2016ode,muehlebach2019dynamical,franca2021flows,hassan2021prox} and on certificates obtained by semidefinite programming \cite{lessard2016iqc,hu2017dissipativity,drori2014pep,taylor2017pep}. We apply this framework to VAMP and present the following three contributions:
\begin{itemize}[leftmargin=*]
\item We describe the circuit associated with VAMP. Its extrinsic update is a mirror resistor, and damping is a lag filter.
\item We show that the loop gain of this circuit explains the oscillation. Its closed-form spectrum can predict when VAMP fails and shows why damping is a heuristic solution.
\item We propose circuit-adaptive damping strategy based on a lag-lead filter designed from the loop gain, often with negative momentum. It outperforms existing damping strategies on ill-conditioned channels.
\end{itemize}

\textit{Notation.} Bold letters denote vectors and matrices, and $\prox_{Rf}(\rr) = \arg\min_{\x} f(\x) + \|\x - \rr\|^2/(2R)$.

\section{The Circuit Analogy of VAMP}
\label{sec:circuit}
Let $\y = \A\x + \w$ with $\A \in \R^{M \times N}$, $\x$ drawn from a separable prior and $\w \sim \mathcal{N}(\bm{0}, \gamma_w^{-1}\I)$. VAMP alternates two modules that exchange an estimate $\rr$ and a precision $\gamma$ (cf. Algorithm~\ref{alg:vamp}). At iteration $k$ the denoiser computes $\x_{1,k} = \bm{g}_1(\rr_{1,k}, \gamma_{1,k})$. It is the posterior mean of $\x$ given $\rr_{1,k} = \x + \mathcal{N}(\bm{0}, (\gamma_{1,k})^{-1}\I)$ with mean slope $\alpha_{1,k}$. The linear minimum mean-square error (LMMSE) module computes $\x_{2,k}$ with slope $\alpha_{2,k}$. Each module returns an extrinsic message by removing the linear part of its input as
\begin{equation}
\gamma_{2,k} = \gamma_{1,k}\Big(\frac{1}{\alpha_{1,k}} - 1\Big),\qquad \rr_{2,k} = \frac{\x_{1,k} - \alpha_{1,k}\rr_{1,k}}{1 - \alpha_{1,k}}.
\label{eq:extrinsic}
\end{equation}
Line~\ref{l:damp} in Algorithm~\ref{alg:vamp} is damping with factor $d \in (0, 1]$. In this work, we write $\tilde\rr_{1,k+1} = \bm{F}(\rr_{1,k})$ for the map of lines~\ref{l:first} to \ref{l:last}, whose fixed points are those of VAMP.

\begin{algorithm}[!t]
\caption{VAMP with damping $d$}
\label{alg:vamp}
\KwIn{$\y, \A, \gamma_w$, denoiser $\bm{g}_1$, damping $d$}
$\rr_{1,0} \gets \bm{0}$, $\gamma_{1,0} \gets \gamma_{\rm init}$\;
\For{$k = 0, 1, \dots$}{
  $\x_{1,k} \gets \bm{g}_1(\rr_{1,k}, \gamma_{1,k})$, \ $\alpha_{1,k} \gets \langle \partial \bm{g}_1/\partial\rr_1\rangle$\; \label{l:first}
  $\gamma_{2,k} \gets \gamma_{1,k}(1/\alpha_{1,k} - 1)$\; \label{l:g2}
  $\rr_{2,k} \gets (\x_{1,k} - \alpha_{1,k} \rr_{1,k})/(1 - \alpha_{1,k})$\;
  $\x_{2,k} \gets (\gamma_w \A^\top\A + \gamma_{2,k}\I)^{-1}(\gamma_w\A^\top\y + \gamma_{2,k}\rr_{2,k})$\;
  $\alpha_{2,k} \gets \tfrac{\gamma_{2,k}}{N}\tr(\gamma_w\A^\top\A + \gamma_{2,k}\I)^{-1}$\;
  $\tilde\gamma_{1,k+1} \gets \gamma_{2,k}(1/\alpha_{2,k} - 1)$\; \label{l:g1}
  $\tilde\rr_{1,k+1} \gets (\x_{2,k} - \alpha_{2,k} \rr_{2,k})/(1 - \alpha_{2,k})$\; \label{l:last}
  $\rr_{1,k+1} \gets d\,\tilde\rr_{1,k+1} + (1-d)\,\rr_{1,k}$\; \label{l:damp}
  $(\gamma_{1,k+1})^{-1/2} \gets d\,(\tilde\gamma_{1,k+1})^{-1/2} + (1-d)\,(\gamma_{1,k})^{-1/2}$\; \label{l:dampg}
}
\end{algorithm}

\subsection{Algorithmic Design via Electric Circuits}
\begin{figure}[!t]
\centering
\includegraphics[width=0.8\columnwidth]{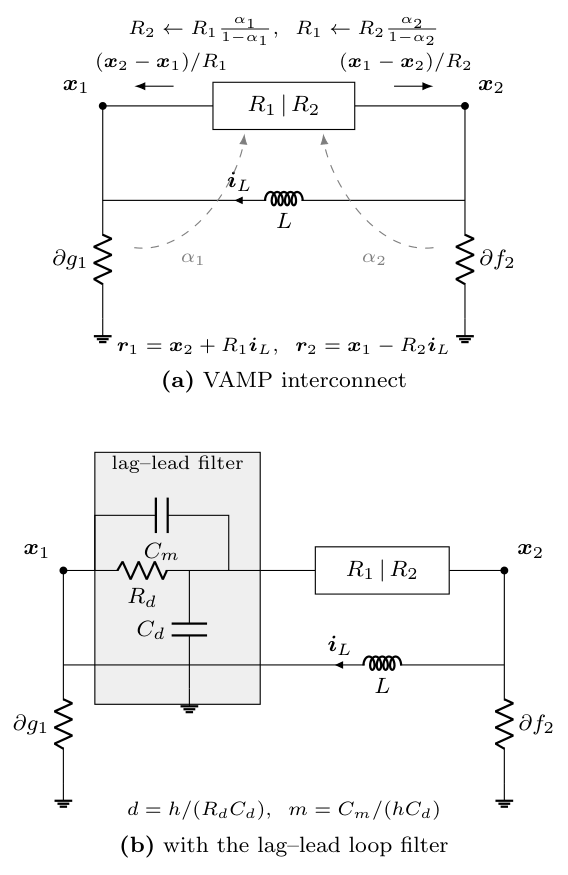}
\caption{(a) The VAMP interconnect. An inductor in parallel with a two-sided resistor, $R_1 = 1/\gamma_1$ and $R_2 = 1/\gamma_2$, joins the denoiser and the likelihood. The precision update makes each side a mirror resistor (dashed arrows). (b) A lag-lead loop filter at the denoiser port. The lag $d$ is set by $R_dC_d$ and the lead $m$ by $C_m$, whose sign is free.}
\label{fig:circuit}
\end{figure}

Building upon \cite{boyd2024circuits}, a convex function $f$ is represented by its subdifferential $\partial f$, a nonlinear resistor whose terminal potential $\x$ and terminal current $\y$ satisfy $\y \in \partial f(\x)$. Our two devices are the denoiser $\partial g_1$, with $g_1 = -\log p(\x)$, and the likelihood $\partial f_2$, with $f_2 = \tfrac{\gamma_w}{2}\|\y - \A\x\|^2$. They are joined by an interconnect. At equilibrium inductors are shorts and capacitors are open, so the interconnect reduces to a wire. Kirchhoff's laws then give $\x_1 = \x_2 = \xs$ with $\bm{0} \in \partial g_1(\xs) + \partial f_2(\xs)$. Every such interconnect is an algorithm with the same fixed point, so the interconnect is the design object. Fig.~\ref{fig:circuit}(a) shows the VAMP interconnect. An inductor $L$ is in parallel with a two-sided resistor. It presents $R_1$ to the denoiser node and $R_2$ to the likelihood node. Kirchhoff's current law at the two nodes gives
\begin{equation}
\x_1 = \prox_{R_1 g_1}(\x_2 + R_1\iL),\qquad \x_2 = \prox_{R_2 f_2}(\x_1 - R_2\iL).
\label{eq:kcl}
\end{equation}
For $R_1 = R_2$ it is the Douglas-Rachford circuit \cite{boyd2024circuits}. For $R_1 \neq R_2$, the branch is a resistor $R_1$ followed by a current mirror of gain $R_1/R_2$, the one non-passive element of the loop.

\begin{proposition}
\label{prop:fixed}
Let $R_1 = 1/\gamma_1$ and $R_2 = 1/\gamma_2$. Integrate the inductor by an Euler step after each device, with step $L/R_j$ after device $j$. With $\rr_{1,k} = \x_{2,k-1} + R_1\bm{i}_{k}$ this gives
\begin{align}
\x_{1,k} &= \prox_{R_1 g_1}(\rr_{1,k}),& \bm{i}_{k+\frac12} &= \bm{i}_{k} + \tfrac{1}{R_1}(\x_{2,k-1} - \x_{1,k}),\label{eq:s1}\\
\x_{2,k} &= \prox_{R_2 f_2}(\rr_{2,k}),& \bm{i}_{k+1} &= \bm{i}_{k+\frac12} + \tfrac{1}{R_2}(\x_{2,k} - \x_{1,k}),\label{eq:s2}
\end{align}
with $\rr_{2,k} = \x_{1,k} - R_2\bm{i}_{k+\frac12}$. The pair $(\rr_{1,k}, \rr_{2,k})$ follows exactly Algorithm~\ref{alg:vamp} at fixed precisions with $d = 1$.
\end{proposition}
\begin{proof}
Let $\y_{1,k} = (\rr_{1,k} - \x_{1,k})/R_1$. The prox identity gives $\x_{1,k} = \bm{g}_1(\rr_{1,k}, \gamma_1)$. At fixed precisions, \eqref{eq:extrinsic} reads $\rr_{2,k} = \x_{1,k} - (\gamma_1/\gamma_2)(\rr_{1,k} - \x_{1,k}) = \x_{1,k} - R_2\y_{1,k}$. In the circuit, $\rr_{2,k} = \x_{1,k} - R_2\bm{i}_{k} - (R_2/R_1)(\x_{2,k-1} - \x_{1,k})$, which equals $\x_{1,k} - (R_2/R_1)(\rr_{1,k} - \x_{1,k}) = \x_{1,k} - R_2\y_{1,k}$. The same computation after device 2 gives $\rr_{1,k+1} = \x_{2,k} - R_1\y_{2,k}$, which is the second extrinsic step.
\end{proof}
\emph{Mirror resistors.} In terms of resistances, the precision updates of lines~\ref{l:g2} and \ref{l:g1} read
\begin{equation}
R_{2,k} = R_{1,k}\,\frac{\alpha_{1,k}}{1 - \alpha_{1,k}},\qquad R_{1,k+1} = R_{2,k}\,\frac{\alpha_{2,k}}{1 - \alpha_{2,k}}.
\label{eq:mirror}
\end{equation}
Each side is reset to the small-signal resistance of the device it faces, so the linear response of the device cancels. This mirror resistor correspond to the Onsager correction.

\subsection{Damping as a Lag Filter}
Line~\ref{l:damp} reads $\rr_{1,k+1} = \rr_{1,k} + d\,(\tilde\rr_{1,k+1} - \rr_{1,k})$. This is the Euler step of $\tau\dot\rr_1 = \tilde\rr_1 - \rr_1$ with $d = h/\tau$. It is a resistor-capacitor low-pass filter at the denoiser port with $\tau = R_dC_d$, the lag of Fig.~\ref{fig:circuit}(b). Its transfer function $d\,z^{-1}/(1 - (1-d)z^{-1})$ has one pole at $1 - d$, and every tuning practice for VAMP chooses this pole.

\begin{figure}[!t]
\centering
\includegraphics[width=0.92\columnwidth]{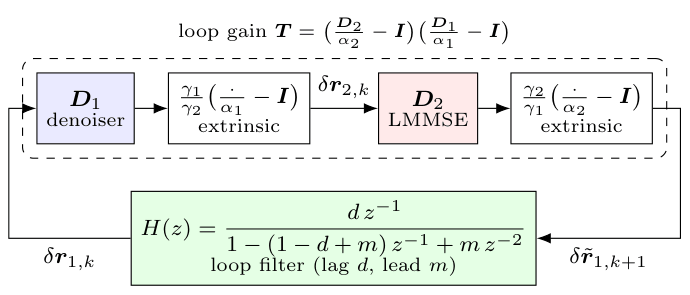}
\caption{The VAMP loop linearized at its fixed point. Each device contributes its small-signal gain. Each extrinsic step subtracts the mean gain through the mirror resistor. The loop closes through the loop filter $H(z)$.}
\label{fig:loop}
\end{figure}

\begin{figure*}[!t]
\centering
\includegraphics[width=0.98\textwidth]{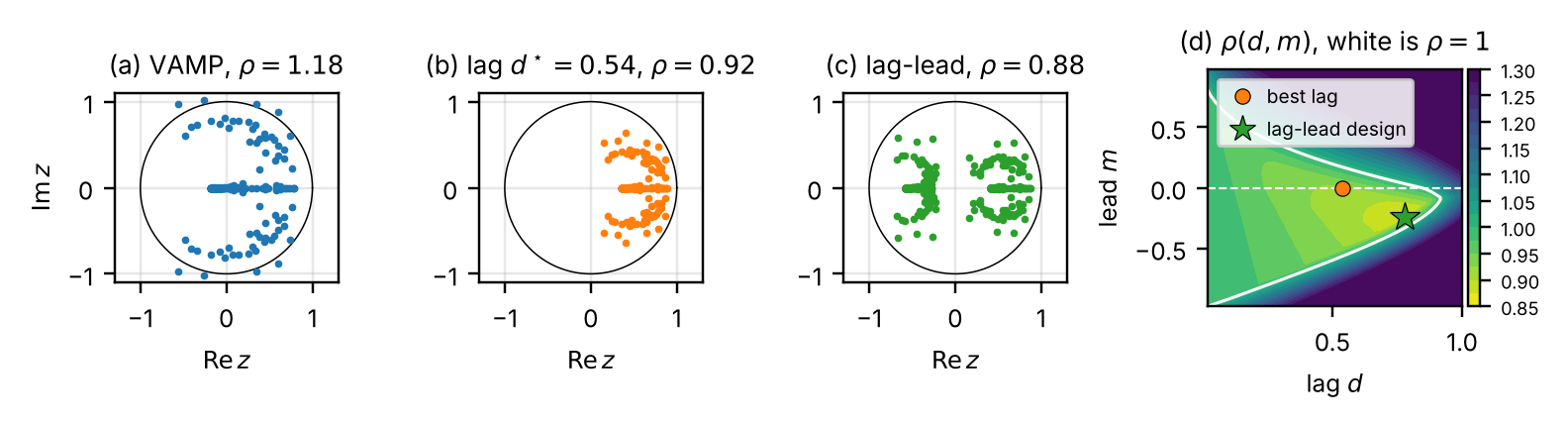}
\vspace{-0.4cm}
\caption{Loop gain at the fixed point of an instance of Section~\ref{sec:exp} with $\kappa = 10^4$. (a) Eigenvalues of $\bm{J}$. (b) Closed-loop poles with the best lag $d^\star$. (c) Closed-loop poles with the lag-lead design \eqref{eq:design}. (d) Spectral radius over $(d, m)$. The dashed line is damping alone.}
\vspace{-0.2cm}
\label{fig:spectra}
\end{figure*}

\section{Circuit-Adaptive Damping}
\label{sec:design}
\subsection{The Loop Gain}
\label{sec:loopgain}
Let $\Dm_1 = \diag(\partial\bm{g}_1/\partial\rr_1)$ be the diagonal signal response of the denoiser. Similarly, let $\Dm_2 = \gamma_2(\gamma_w\A^\top\A + \gamma_2\I)^{-1}$ be that of the LMMSE module. Fig.~\ref{fig:loop} shows the linearized loop.
\begin{proposition}
\label{prop:T}
Linearize Algorithm~\ref{alg:vamp} at a fixed point. A perturbation $\dr_{1,k}$ of the denoiser message propagates through lines~\ref{l:first} to \ref{l:last} as $\delta\tilde\rr_{1,k+1} = \bm{J}\,\dr_{1,k}$, with
\begin{equation}
\bm{J} = \T + \tfrac{1}{N}\,\bm{w}\bm{h}^\top,~\textrm{with}~ \T = \Big(\frac{\Dm_2}{\alpha_2} - \I\Big)\Big(\frac{\Dm_1}{\alpha_1} - \I\Big).
\label{eq:T}
\end{equation}
Here $h_i = \partial[\Dm_1]_{ii}/\partial[\rr_1]_i$ is the curvature of the denoiser, and
\begin{equation*}
\begin{aligned}
\bm{w} = {}&\frac{\Dm_2 - \alpha_2\I}{1-\alpha_2}\,\frac{\x_1 - \rr_1}{(1-\alpha_1)^2}\\
&~~- \frac{\gamma_1}{\alpha_1^2}\Big(\frac{\bm{Q}(\rr_2 - \x_2)}{1-\alpha_2} + \frac{\beta_2(\x_2 - \rr_2)}{(1-\alpha_2)^2}\Big),
\end{aligned}
\end{equation*}
with $\bm{Q} = (\gamma_w\A^\top\A + \gamma_2\I)^{-1}$ and $\beta_2 = \tfrac{\gamma_w}{N}\tr(\bm{Q}\A^\top\A\bm{Q})$. Through the damped loop the perturbation propagates as $\dr_{1,k+1} = (d\bm{J} + (1-d)\I)\,\dr_{1,k}$.
\end{proposition}
\begin{proof}
Differentiate line~\ref{l:first} to line~\ref{l:last}. The mean slope moves by $\delta\alpha_1 = \bm{h}^\top\dr_1/N$, and the precision by $\delta\gamma_2 = -(\gamma_1/\alpha_1^2)\,\delta\alpha_1$. The first extrinsic step gives $\dr_2 = (\Dm_1 - \alpha_1\I)\dr_1/(1-\alpha_1) + (\x_1 - \rr_1)\,\delta\alpha_1/(1-\alpha_1)^2$. The LMMSE module gives $\delta\x_2 = \Dm_2\dr_2 + \bm{Q}(\rr_2 - \x_2)\delta\gamma_2$ and $\delta\alpha_2 = \beta_2\delta\gamma_2$. The second extrinsic step then gives \eqref{eq:T}. At a fixed point $(1-\alpha_1)(1-\alpha_2) = \alpha_1\alpha_2$, which turns the product of the two responses into $\T$.
\end{proof}

The rank-one term $\tfrac{1}{N}\bm{w}\bm{h}^\top$ in \eqref{eq:T} comes from the change of the mean slope $\alpha_1$. It moves a single collective mode, in which all coordinates shift together. Each factor of $\T$ is a device response centred by its mirror resistor. The denoiser factor equals $-1$ off the support where the denoiser outputs zero, and is large and positive on it. The LMMSE factor equals $-1$ on the directions that the measurements determine and is positive on those hidden by the noise. On an ill-conditioned matrix, this factor varies strongly across directions. Its product with the large on-support gain can then have eigenvalues of modulus above one.

Fig.~\ref{fig:spectra}(a) shows the spectrum of $\bm{J}$ on an instance with a condition number $\kappa = 10^4$. Complex modes at $-0.11 \pm 1.17j$ and $-0.57 \pm 0.97j$ lie outside the unit circle, and undamped VAMP does not converge. For an eigenvalue in the left half-plane, the component of a perturbation along its eigenvector nearly reverses at every iteration. The two modules then over-correct each other in turn. Over the instances of Section~\ref{sec:exp} whose fixed point is reached, the test $\rho(\bm{J}) < 1$ predicts in most cases whether undamped VAMP converges. The same test on $\T$ alone is much less accurate, so the rank-one term matters.

Damping maps every eigenvalue by $\lambda \mapsto d\lambda + 1 - d$, which moves it on a straight line toward $+1$ (Fig.~\ref{fig:spectra}(b)). The modes at $-0.11 \pm 1.17j$ enter the unit circle only for $d < 0.85$. A smaller $d$ also pulls the modes near $0.74 \pm 0.60j$ toward $+1$, where they decay slowly. The best factor $d^\star = 0.54$ still leaves $\rho = 0.92$. A single factor cannot serve both groups of modes, which is the limit of damping.

\subsection{Lag-Lead Loop Filter Design}
\label{sec:filter}
In control engineering, an oscillating loop is stabilized by a lead compensator. In our circuit, it is a capacitor whose current is proportional to the rate of change of the message. Fig.~\ref{fig:circuit}(b) adds this capacitor, and line~\ref{l:damp} becomes
\begin{equation}
\rr_{1,k+1} = d\,\tilde\rr_{1,k+1} + (1 - d)\,\rr_{1,k} + m\,(\rr_{1,k} - \rr_{1,k-1}).
\label{eq:update}
\end{equation}
Here $d = h/(R_dC_d)$ is the lag and $m = C_m/(hC_d)$ is the lead, as in the heavy-ball method \cite{polyak1964}. The loop filter of Fig.~\ref{fig:loop} becomes $H(z) = d\,z^{-1}/(1 - (1-d+m)z^{-1} + m z^{-2})$. The lead $m(1 - z^{-1})$ has gain $2|m\sin(\omega/2)|$, zero at $\omega = 0$ and largest at $\omega = \pi$. It leaves the slow modes alone and acts on the oscillating ones.

For an eigenvalue $\lambda$ of $\T$, the closed loop of Fig.~\ref{fig:loop} has the two poles
\begin{equation}
z_\pm(\lambda, d, m) = \tfrac12\Big[\mu \pm \sqrt{\mu^2 - 4m}\Big],\qquad \mu = d\lambda + 1 - d + m.
\label{eq:roots}
\end{equation}
The mode decays at the rate $z(\lambda, d, m) = \max\{|z_+|, |z_-|\}$. The design problem is then
\begin{equation}
(d, m) = \arg\min_{0 < d \le 1,\ |m| < 1}\;\max_{\lambda \in \Lambda}\,z(\lambda, d, m),
\label{eq:design}
\end{equation}
where $\Lambda$ is the set of stabilizable eigenvalues of the loop gain. An eigenvalue $\lambda$ is stabilizable when some admissible $(d, m)$ gives $z(\lambda, d, m) < 1$.
A convex combination needs $0 < d \le 1$, and $z_+ z_- = m$ forces $|m| < 1$. Without the restriction to $\Lambda$, a mode no filter can stabilize would drive the design to $d \to 0$ and freeze the iteration. We solve \eqref{eq:design} on a grid of step $0.02$.

In Fig.~\ref{fig:spectra}(d) the stable region tilts toward negative $m$, and the design reaches $\rho = 0.88$ with $m = -0.24$ (Fig.~\ref{fig:spectra}(c)). Accelerated gradient descent uses a positive lead, because the slow modes of a gradient flow sit near $+1$ \cite{boyd2024circuits,su2016ode}. Here the modes to control sit in the left half-plane, and the design reverses the sign of the capacitor. For $\kappa \ge 10^3$ the design selects a negative lead in about 40\% of the redesigns.

\begin{figure}[!b]
\centering
\includegraphics[width=\columnwidth]{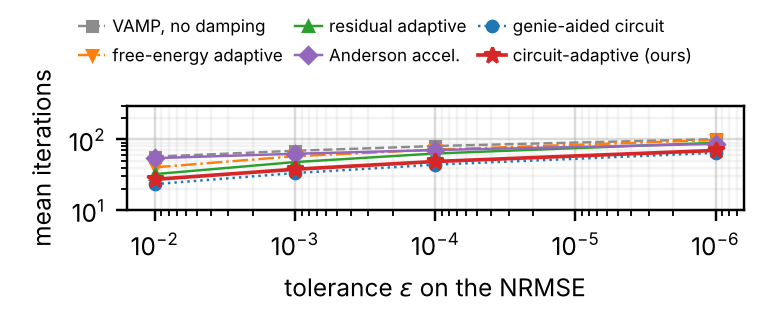}
\vspace{-0.7cm}
\caption{Mean iterations over the instances of Fig.~\ref{fig:sweep} versus the tolerance $\epsilon$.}
\label{fig:tol}
\end{figure}

VAMP with \emph{circuit-adaptive} damping is Algorithm~\ref{alg:vamp} with line~\ref{l:damp} replaced by \eqref{eq:update}. It starts undamped with $(d, m) = (1, 0)$. Every $K$ iterations it redesigns $(d, m)$ by \eqref{eq:design} from $\T$ at the current state, since the fixed point is unknown. Both factors of $\T$ come at no extra cost, i.e., from the denoiser slopes and the eigendecomposition of $\A^\top\A$ used by the LMMSE module. We use $\T$ rather than $\bm{J}$, because the rank-one term of \eqref{eq:T} understates the damping needed during the transient regime. Designing from $\bm{J}$ instead of $\T$ fails on many more instances. At large $N$, the extreme eigenvalues that matter in \eqref{eq:design} can be computed by the Arnoldi method \cite{sorensen1992implicit} from products of $\T$ with vectors.

\section{Experiments}
\label{sec:exp}
Every matrix is the real form of a multipath channel with $r/2$ paths between linear arrays of 75 and 150 antennas, with $N = 300$ and $M = 150$. Its singular value decomposition gives structured $\bm{U}$ and $\bm{V}$. We keep them and set the $r$ nonzero singular values from 1 to $1/\kappa$ geometrically. Unless stated otherwise, $r = 150$, the Bernoulli-Gaussian prior has sparsity $0.1$ and the signal-to-noise ratio is 30~dB. Every point averages 1000 Monte Carlo instances.

\begin{figure}[!t]
\centering
\includegraphics[width=\columnwidth]{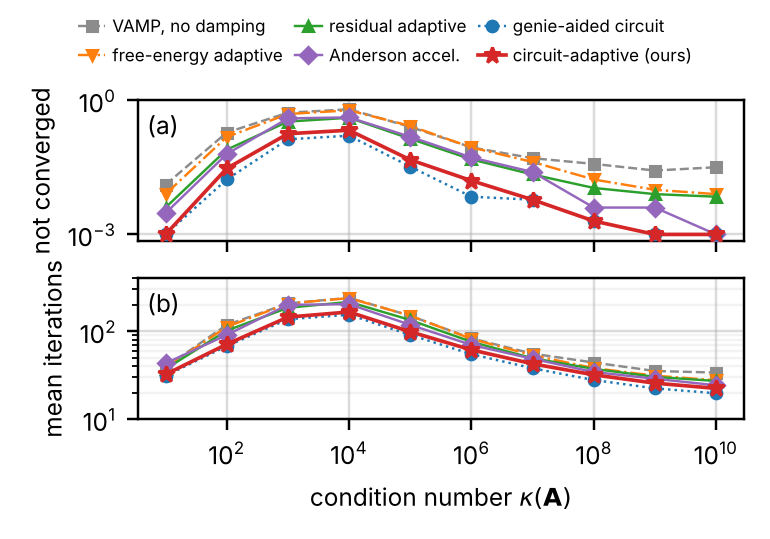}
\vspace{-0.35cm}
\caption{Condition-number sweep at $\epsilon = 10^{-6}$. (a) Fraction not converged, with zero drawn at $10^{-3}$. (b) Mean iterations.}
\label{fig:sweep}
\end{figure}

Let $e_k = \|\x_{2,k} - \x\|/\|\x\|$ be the normalized root mean-square error (NRMSE) at iteration $k$. Let $e^\star$ be its value at the VAMP fixed point, taken from the run with the smallest fixed-point residual. A run has converged to tolerance $\epsilon$ at the first iteration after which $|e_k - e^\star| \le \epsilon\, e^\star$.

We compare circuit-adaptive VAMP, which redesigns its filter every $K = 15$ iterations, against undamped VAMP and three practical strategies. Residual adaptive damping halves $d$ when the residual exceeds its maximum over the last five iterations and grows it by ten percent otherwise. Free-energy adaptive damping combines the expectation-consistent (EC) free energy of \cite{fletcher2016ec} as its cost with the step-size rule of \cite{vila2015damping}. It halves $d$ and retries the step when the cost exceeds its last four accepted values. Anderson acceleration extrapolates from the last four iterates and averages the result with its image under $\bm{F}$ \cite{walker2011anderson}. Genie-aided circuit damping runs the fixed filter \eqref{eq:update} for every $d \in \{0.1, \dots, 1\}$ and $m \in \{-0.5, \dots, 0.3\}$, from the start and after our 15 undamped iterations. It retains the fastest of these 180 runs in hindsight, making it a genie-aided benchmark rather than a practical damping strategy.

Fig.~\ref{fig:tol} shows that our circuit-adaptive VAMP is the fastest practical strategy at every tolerance. At $\epsilon = 10^{-6}$ it needs 19\% fewer iterations than Anderson acceleration, 22\% fewer than residual adaptive damping and 29\% fewer than free-energy adaptive damping. It needs only 8\% more than the genie, which uses 180 runs instead of one.

Fig.~\ref{fig:sweep} sweeps $\kappa$ from $10$ to $10^{10}$. Circuit-adaptive VAMP converges on 95.0\% of the instances, against 88.8\% for Anderson acceleration, 89.3\% for residual adaptive damping, 83.0\% for free-energy adaptive damping and 96.4\% for the genie. On the hardest decades, $\kappa = 10^3$ and $10^4$, it converges on 80.2\% of the instances, against at most 62.6\%.

\begin{figure}[!t]
\centering
\includegraphics[width=\columnwidth]{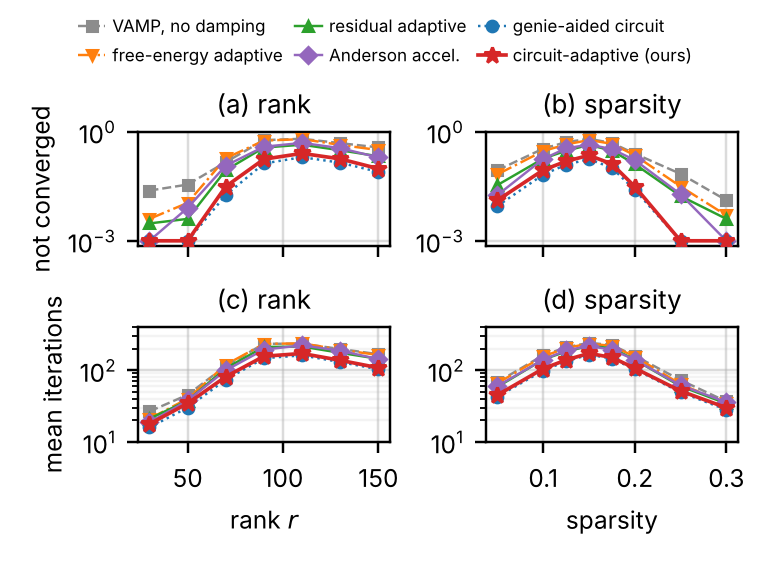}
\vspace{-0.35cm}
\caption{Rank and sparsity sweeps at $\kappa = 300$ and $\epsilon = 10^{-6}$. (a), (c) Rank sweep at sparsity $0.1$. (b), (d) Sparsity sweep at rank $150$.}
\label{fig:lowrank}
\end{figure}

\textbf{Noise.} For signal-to-noise ratios from 10 to 50~dB and $\kappa \in \{10^3, 10^4, 10^5\}$, our method converges on 90.3\% of the instances. Residual adaptive damping converges on 82.4\%, Anderson acceleration on 79.3\% and free-energy adaptive damping on 72.8\%.

\textbf{Rank and sparsity.} Fig.~\ref{fig:lowrank}(a) and (c) sweep the rank $r$ at 10\%s parsity $0.1$. Our method converges on 89.5\% of the instances, against 78.0\% for Anderson acceleration, 79.8\% for residual adaptive damping, 69.6\% for free-energy adaptive damping and 91.9\% for the genie. Fig.~\ref{fig:lowrank}(b) and (d) sweep the sparsity at full rank. Our method converges on 92.0\% of the instances, against 81.3\% for Anderson acceleration, 82.6\% for residual adaptive damping, 73.6\% for free-energy adaptive damping and 93.8\% for the genie.

\section{Conclusion}
VAMP admits a circuit interpretation in which non-convergence corresponds to a loop instability revealed by a closed-form loop gain. To suppress the oscillating modes that damping cannot isolate, we introduce a lag-lead loop filter whose lead component is often negative. Designed from the loop gain at the current state, it outperforms existing damping strategies on ill-conditioned matrices.

\clearpage
\bibliographystyle{IEEEtran}
\bibliography{refs}

@article{donoho2009amp,
  author={Donoho, David L. and Maleki, Arian and Montanari, Andrea},
  title={Message-passing algorithms for compressed sensing},
  journal={Proc. Nat. Acad. Sci.}, volume={106}, number={45}, pages={18914--18919}, year={2009}}

@inproceedings{rangan2011gamp,
  author={Rangan, Sundeep},
  title={Generalized approximate message passing for estimation with random linear mixing},
  booktitle={Proc. IEEE Int. Symp. Inf. Theory}, pages={2168--2172}, year={2011}}

@article{rangan2019vamp,
  author={Rangan, Sundeep and Schniter, Philip and Fletcher, Alyson K.},
  title={Vector approximate message passing},
  journal={IEEE Trans. Inf. Theory}, volume={65}, number={10}, pages={6664--6684}, year={2019}}

@article{ma2017oamp,
  author={Ma, Junjie and Ping, Li},
  title={Orthogonal {AMP}},
  journal={IEEE Access}, volume={5}, pages={2020--2033}, year={2017}}

@article{takeuchi2020ep,
  author={Takeuchi, Keigo},
  title={Rigorous dynamics of expectation-propagation-based signal recovery from unitarily invariant measurements},
  journal={IEEE Trans. Inf. Theory}, volume={66}, number={1}, pages={368--386}, year={2020}}

@article{rangan2019convergence,
  author={Rangan, Sundeep and Schniter, Philip and Fletcher, Alyson K. and Sarkar, Subrata},
  title={On the convergence of approximate message passing with arbitrary matrices},
  journal={IEEE Trans. Inf. Theory}, volume={65}, number={9}, pages={5339--5351}, year={2019}}

@inproceedings{vila2015damping,
  author={Vila, Jeremy and Schniter, Philip and Rangan, Sundeep and Krzakala, Florent and Zdeborov{\'a}, Lenka},
  title={Adaptive damping and mean removal for the generalized approximate message passing algorithm},
  booktitle={Proc. IEEE Int. Conf. Acoust., Speech, Signal Process.}, pages={2021--2025}, year={2015}}

@article{liu2022mamp,
  author={Liu, Lei and Huang, Shunqi and Kurkoski, Brian M.},
  title={Memory {AMP}},
  journal={IEEE Trans. Inf. Theory}, volume={68}, number={12}, pages={8015--8039}, year={2022}}

@article{walker2011anderson,
  author={Walker, Homer F. and Ni, Peng},
  title={Anderson acceleration for fixed-point iterations},
  journal={SIAM J. Numer. Anal.}, volume={49}, number={4}, pages={1715--1735}, year={2011}}

@article{polyak1964,
  author={Polyak, Boris T.},
  title={Some methods of speeding up the convergence of iteration methods},
  journal={USSR Comput. Math. Math. Phys.}, volume={4}, number={5}, pages={1--17}, year={1964}}

@article{su2016ode,
  author={Su, Weijie and Boyd, Stephen and Cand{\`e}s, Emmanuel J.},
  title={A differential equation for modeling {N}esterov's accelerated gradient method: Theory and insights},
  journal={J. Mach. Learn. Res.}, volume={17}, number={153}, pages={1--43}, year={2016}}

@article{lessard2016iqc,
  author={Lessard, Laurent and Recht, Benjamin and Packard, Andrew},
  title={Analysis and design of optimization algorithms via integral quadratic constraints},
  journal={SIAM J. Optim.}, volume={26}, number={1}, pages={57--95}, year={2016}}

@article{drori2014pep,
  author={Drori, Yoel and Teboulle, Marc},
  title={Performance of first-order methods for smooth convex minimization: A novel approach},
  journal={Math. Program.}, volume={145}, pages={451--482}, year={2014}}

@article{taylor2017pep,
  author={Taylor, Adrien B. and Hendrickx, Julien M. and Glineur, Fran{\c{c}}ois},
  title={Smooth strongly convex interpolation and exact worst-case performance of first-order methods},
  journal={Math. Program.}, volume={161}, pages={307--345}, year={2017}}

@inproceedings{boyd2024circuits,
  author={Boyd, Stephen P. and Parshakova, Tetiana and Ryu, Ernest K. and Suh, Jaewook J.},
  title={Optimization algorithm design via electric circuits},
  booktitle={Advances in Neural Information Processing Systems}, volume={37}, year={2024}, note={arXiv:2411.02573}}

@article{franca2021flows,
  author={Fran{\c{c}}a, Guilherme and Robinson, Daniel P. and Vidal, Ren{\'e}},
  title={Gradient flows and proximal splitting methods: A unified view on accelerated and stochastic optimization},
  journal={Phys. Rev. E}, volume={103}, number={5}, pages={053304}, year={2021}}

@article{hassan2021prox,
  author={Hassan-Moghaddam, Sepideh and Jovanovi{\'c}, Mihailo R.},
  title={Proximal gradient flow and {D}ouglas-{R}achford splitting dynamics: Global exponential stability via integral quadratic constraints},
  journal={Automatica}, volume={123}, pages={109311}, year={2021}}

@inproceedings{muehlebach2019dynamical,
  author={Muehlebach, Michael and Jordan, Michael I.},
  title={A dynamical systems perspective on {N}esterov acceleration},
  booktitle={Proc. Int. Conf. Mach. Learn.}, pages={4656--4662}, year={2019}}

@inproceedings{hu2017dissipativity,
  author={Hu, Bin and Lessard, Laurent},
  title={Dissipativity theory for {N}esterov's accelerated method},
  booktitle={Proc. Int. Conf. Mach. Learn.}, pages={1549--1557}, year={2017}}

@article{sorensen1992implicit,
  author={Sorensen, Danny C.},
  title={Implicit application of polynomial filters in a $k$-step {A}rnoldi method},
  journal={SIAM J. Matrix Anal. Appl.}, volume={13}, number={1}, pages={357--385}, year={1992}}

@inproceedings{fletcher2016ec,
  author    = {Fletcher, Alyson K. and Sahraee-Ardakan, Mojtaba and Rangan, Sundeep and Schniter, Philip},
  title     = {Expectation consistent approximate inference: {G}eneralizations and convergence},
  booktitle = {Proc. IEEE Int. Symp. Inf. Theory},
  pages     = {190--194},
  year      = {2016}
}

\end{document}